\documentclass[letterpaper,compsoc,twoside]{IEEEtran}
\usepackage{fixltx2e} 
\usepackage{cmap} 
\usepackage{ifthen}
\usepackage[T1]{fontenc}
\usepackage[utf8]{inputenc}

\usepackage[font={small,it},labelfont=bf]{caption}
\usepackage{float}

\pdfoutput=1
\usepackage{scipy}
\makeatletter
\def\PY@reset{\let\PY@it=\relax \let\PY@bf=\relax%
    \let\PY@ul=\relax \let\PY@tc=\relax%
    \let\PY@bc=\relax \let\PY@ff=\relax}
\def\PY@tok#1{\csname PY@tok@#1\endcsname}
\def\PY@toks#1+{\ifx\relax#1\empty\else%
    \PY@tok{#1}\expandafter\PY@toks\fi}
\def\PY@do#1{\PY@bc{\PY@tc{\PY@ul{%
    \PY@it{\PY@bf{\PY@ff{#1}}}}}}}
\def\PY#1#2{\PY@reset\PY@toks#1+\relax+\PY@do{#2}}

\expandafter\def\csname PY@tok@gd\endcsname{\def\PY@tc##1{\textcolor[rgb]{0.63,0.00,0.00}{##1}}}
\expandafter\def\csname PY@tok@gu\endcsname{\let\PY@bf=\textbf\def\PY@tc##1{\textcolor[rgb]{0.50,0.00,0.50}{##1}}}
\expandafter\def\csname PY@tok@gt\endcsname{\def\PY@tc##1{\textcolor[rgb]{0.00,0.27,0.87}{##1}}}
\expandafter\def\csname PY@tok@gs\endcsname{\let\PY@bf=\textbf}
\expandafter\def\csname PY@tok@gr\endcsname{\def\PY@tc##1{\textcolor[rgb]{1.00,0.00,0.00}{##1}}}
\expandafter\def\csname PY@tok@cm\endcsname{\let\PY@it=\textit\def\PY@tc##1{\textcolor[rgb]{0.25,0.50,0.56}{##1}}}
\expandafter\def\csname PY@tok@vg\endcsname{\def\PY@tc##1{\textcolor[rgb]{0.73,0.38,0.84}{##1}}}
\expandafter\def\csname PY@tok@m\endcsname{\def\PY@tc##1{\textcolor[rgb]{0.13,0.50,0.31}{##1}}}
\expandafter\def\csname PY@tok@mh\endcsname{\def\PY@tc##1{\textcolor[rgb]{0.13,0.50,0.31}{##1}}}
\expandafter\def\csname PY@tok@cs\endcsname{\def\PY@tc##1{\textcolor[rgb]{0.25,0.50,0.56}{##1}}\def\PY@bc##1{\setlength{\fboxsep}{0pt}\colorbox[rgb]{1.00,0.94,0.94}{\strut ##1}}}
\expandafter\def\csname PY@tok@ge\endcsname{\let\PY@it=\textit}
\expandafter\def\csname PY@tok@vc\endcsname{\def\PY@tc##1{\textcolor[rgb]{0.73,0.38,0.84}{##1}}}
\expandafter\def\csname PY@tok@il\endcsname{\def\PY@tc##1{\textcolor[rgb]{0.13,0.50,0.31}{##1}}}
\expandafter\def\csname PY@tok@go\endcsname{\def\PY@tc##1{\textcolor[rgb]{0.20,0.20,0.20}{##1}}}
\expandafter\def\csname PY@tok@cp\endcsname{\def\PY@tc##1{\textcolor[rgb]{0.00,0.44,0.13}{##1}}}
\expandafter\def\csname PY@tok@gi\endcsname{\def\PY@tc##1{\textcolor[rgb]{0.00,0.63,0.00}{##1}}}
\expandafter\def\csname PY@tok@gh\endcsname{\let\PY@bf=\textbf\def\PY@tc##1{\textcolor[rgb]{0.00,0.00,0.50}{##1}}}
\expandafter\def\csname PY@tok@ni\endcsname{\let\PY@bf=\textbf\def\PY@tc##1{\textcolor[rgb]{0.84,0.33,0.22}{##1}}}
\expandafter\def\csname PY@tok@nl\endcsname{\let\PY@bf=\textbf\def\PY@tc##1{\textcolor[rgb]{0.00,0.13,0.44}{##1}}}
\expandafter\def\csname PY@tok@nn\endcsname{\let\PY@bf=\textbf\def\PY@tc##1{\textcolor[rgb]{0.05,0.52,0.71}{##1}}}
\expandafter\def\csname PY@tok@no\endcsname{\def\PY@tc##1{\textcolor[rgb]{0.38,0.68,0.84}{##1}}}
\expandafter\def\csname PY@tok@na\endcsname{\def\PY@tc##1{\textcolor[rgb]{0.25,0.44,0.63}{##1}}}
\expandafter\def\csname PY@tok@nb\endcsname{\def\PY@tc##1{\textcolor[rgb]{0.00,0.44,0.13}{##1}}}
\expandafter\def\csname PY@tok@nc\endcsname{\let\PY@bf=\textbf\def\PY@tc##1{\textcolor[rgb]{0.05,0.52,0.71}{##1}}}
\expandafter\def\csname PY@tok@nd\endcsname{\let\PY@bf=\textbf\def\PY@tc##1{\textcolor[rgb]{0.33,0.33,0.33}{##1}}}
\expandafter\def\csname PY@tok@ne\endcsname{\def\PY@tc##1{\textcolor[rgb]{0.00,0.44,0.13}{##1}}}
\expandafter\def\csname PY@tok@nf\endcsname{\def\PY@tc##1{\textcolor[rgb]{0.02,0.16,0.49}{##1}}}
\expandafter\def\csname PY@tok@si\endcsname{\let\PY@it=\textit\def\PY@tc##1{\textcolor[rgb]{0.44,0.63,0.82}{##1}}}
\expandafter\def\csname PY@tok@s2\endcsname{\def\PY@tc##1{\textcolor[rgb]{0.25,0.44,0.63}{##1}}}
\expandafter\def\csname PY@tok@vi\endcsname{\def\PY@tc##1{\textcolor[rgb]{0.73,0.38,0.84}{##1}}}
\expandafter\def\csname PY@tok@nt\endcsname{\let\PY@bf=\textbf\def\PY@tc##1{\textcolor[rgb]{0.02,0.16,0.45}{##1}}}
\expandafter\def\csname PY@tok@nv\endcsname{\def\PY@tc##1{\textcolor[rgb]{0.73,0.38,0.84}{##1}}}
\expandafter\def\csname PY@tok@s1\endcsname{\def\PY@tc##1{\textcolor[rgb]{0.25,0.44,0.63}{##1}}}
\expandafter\def\csname PY@tok@gp\endcsname{\let\PY@bf=\textbf\def\PY@tc##1{\textcolor[rgb]{0.78,0.36,0.04}{##1}}}
\expandafter\def\csname PY@tok@sh\endcsname{\def\PY@tc##1{\textcolor[rgb]{0.25,0.44,0.63}{##1}}}
\expandafter\def\csname PY@tok@ow\endcsname{\let\PY@bf=\textbf\def\PY@tc##1{\textcolor[rgb]{0.00,0.44,0.13}{##1}}}
\expandafter\def\csname PY@tok@sx\endcsname{\def\PY@tc##1{\textcolor[rgb]{0.78,0.36,0.04}{##1}}}
\expandafter\def\csname PY@tok@bp\endcsname{\def\PY@tc##1{\textcolor[rgb]{0.00,0.44,0.13}{##1}}}
\expandafter\def\csname PY@tok@c1\endcsname{\let\PY@it=\textit\def\PY@tc##1{\textcolor[rgb]{0.25,0.50,0.56}{##1}}}
\expandafter\def\csname PY@tok@kc\endcsname{\let\PY@bf=\textbf\def\PY@tc##1{\textcolor[rgb]{0.00,0.44,0.13}{##1}}}
\expandafter\def\csname PY@tok@c\endcsname{\let\PY@it=\textit\def\PY@tc##1{\textcolor[rgb]{0.25,0.50,0.56}{##1}}}
\expandafter\def\csname PY@tok@mf\endcsname{\def\PY@tc##1{\textcolor[rgb]{0.13,0.50,0.31}{##1}}}
\expandafter\def\csname PY@tok@err\endcsname{\def\PY@bc##1{\setlength{\fboxsep}{0pt}\fcolorbox[rgb]{1.00,0.00,0.00}{1,1,1}{\strut ##1}}}
\expandafter\def\csname PY@tok@kd\endcsname{\let\PY@bf=\textbf\def\PY@tc##1{\textcolor[rgb]{0.00,0.44,0.13}{##1}}}
\expandafter\def\csname PY@tok@ss\endcsname{\def\PY@tc##1{\textcolor[rgb]{0.32,0.47,0.09}{##1}}}
\expandafter\def\csname PY@tok@sr\endcsname{\def\PY@tc##1{\textcolor[rgb]{0.14,0.33,0.53}{##1}}}
\expandafter\def\csname PY@tok@mo\endcsname{\def\PY@tc##1{\textcolor[rgb]{0.13,0.50,0.31}{##1}}}
\expandafter\def\csname PY@tok@mi\endcsname{\def\PY@tc##1{\textcolor[rgb]{0.13,0.50,0.31}{##1}}}
\expandafter\def\csname PY@tok@kn\endcsname{\let\PY@bf=\textbf\def\PY@tc##1{\textcolor[rgb]{0.00,0.44,0.13}{##1}}}
\expandafter\def\csname PY@tok@o\endcsname{\def\PY@tc##1{\textcolor[rgb]{0.40,0.40,0.40}{##1}}}
\expandafter\def\csname PY@tok@kr\endcsname{\let\PY@bf=\textbf\def\PY@tc##1{\textcolor[rgb]{0.00,0.44,0.13}{##1}}}
\expandafter\def\csname PY@tok@s\endcsname{\def\PY@tc##1{\textcolor[rgb]{0.25,0.44,0.63}{##1}}}
\expandafter\def\csname PY@tok@kp\endcsname{\def\PY@tc##1{\textcolor[rgb]{0.00,0.44,0.13}{##1}}}
\expandafter\def\csname PY@tok@w\endcsname{\def\PY@tc##1{\textcolor[rgb]{0.73,0.73,0.73}{##1}}}
\expandafter\def\csname PY@tok@kt\endcsname{\def\PY@tc##1{\textcolor[rgb]{0.56,0.13,0.00}{##1}}}
\expandafter\def\csname PY@tok@sc\endcsname{\def\PY@tc##1{\textcolor[rgb]{0.25,0.44,0.63}{##1}}}
\expandafter\def\csname PY@tok@sb\endcsname{\def\PY@tc##1{\textcolor[rgb]{0.25,0.44,0.63}{##1}}}
\expandafter\def\csname PY@tok@k\endcsname{\let\PY@bf=\textbf\def\PY@tc##1{\textcolor[rgb]{0.00,0.44,0.13}{##1}}}
\expandafter\def\csname PY@tok@se\endcsname{\let\PY@bf=\textbf\def\PY@tc##1{\textcolor[rgb]{0.25,0.44,0.63}{##1}}}
\expandafter\def\csname PY@tok@sd\endcsname{\let\PY@it=\textit\def\PY@tc##1{\textcolor[rgb]{0.25,0.44,0.63}{##1}}}

\def\PYZus{\char`\_}
\def\PYZob{\char`\{}
\def\PYZcb{\char`\}}

\def\PYZsh{\char`\#}

\def\PYZsq{\char`\'}
\def\PYZdq{\char`\"}

\makeatother

\providecommand*{\DUprovidelength}[2]{
  \ifthenelse{\isundefined{#1}}{\newlength{#1}\setlength{#1}{#2}}{}
}

\providecommand*{\DUrole}[2]{%
  \ifcsname DUrole#1\endcsname%
    \csname DUrole#1\endcsname{#2}%
  \else
    \ifcsname docutilsrole#1\endcsname%
      \csname docutilsrole#1\endcsname{#2}%
    \else%
      #2%
    \fi%
  \fi%
}

\DUprovidelength{\DUlineblockindent}{2.5em}
\ifthenelse{\isundefined{\DUlineblock}}{
  \newenvironment{DUlineblock}[1]{%
    \list{}{\setlength{\partopsep}{\parskip}
            \addtolength{\partopsep}{\baselineskip}
            \setlength{\topsep}{0pt}
            \setlength{\itemsep}{0.15\baselineskip}
            \setlength{\parsep}{0pt}
            \setlength{\leftmargin}{#1}}
    \raggedright
  }
  {\endlist}
}{}

\ifthenelse{\isundefined{\hypersetup}}{
  \usepackage[colorlinks=true,linkcolor=blue,urlcolor=blue]{hyperref}
  \urlstyle{same} 
}{}

\begin{document}
\newcounter{footnotecounter}\title{Mining online social networks with Python to study urban mobility}\author{Antònia Tugores$^{\setcounter{footnotecounter}{1}\fnsymbol{footnotecounter}\setcounter{footnotecounter}{2}\fnsymbol{footnotecounter}}$%
          \setcounter{footnotecounter}{1}\thanks{\fnsymbol{footnotecounter} %
          Corresponding author: \protect\href{mailto:antonia@ifisc.uib-csic.es}{antonia@ifisc.uib-csic.es}}\setcounter{footnotecounter}{2}\thanks{\fnsymbol{footnotecounter} Institute for Cross-Disciplinary Physics and Complex Systems, IFISC (UIB-CSIC)}, Pere Colet$^{\setcounter{footnotecounter}{2}\fnsymbol{footnotecounter}}$\thanks{%

          \noindent%
          Copyright\,\copyright\,2014 Antònia Tugores et al. This is an open-access article distributed under the terms of the Creative Commons Attribution License, which permits unrestricted use, distribution, and reproduction in any medium, provided the original author and source are credited. http://creativecommons.org/licenses/by/3.0/%
        }}\maketitle
          \renewcommand{\leftmark}{PROC. OF THE 6th EUR. CONF. ON PYTHON IN SCIENCE (EUROSCIPY 2013)}
          \renewcommand{\rightmark}{MINING ONLINE SOCIAL NETWORKS WITH PYTHON TO STUDY URBAN MOBILITY}

\setcounter{page}{21}
\newcommand*{\docutilsroleref}{\ref}
\newcommand*{\docutilsrolelabel}{\label}
\AtEndDocument{\cleardoublepage}
\begin{abstract}On-line  social networks have grown quickly over the last few years and
nowadays many people use them frequently.  Furthermore the emergence of
smartphones allows to access these networks any time from any physical location.
Among the social networks, Twitter offers a particularly large set of data publicly available.
Here we discuss the procedure to mine this data and store it in
distributed databases using Python scripts. We also illustrate how geolocated tweets can be used
to study the mobility of people in urban areas.\end{abstract}

\section{Introduction%
  \label{introduction}%
}

Although online social networks appeared already in the later nineties, their growth
has taken place mainly in the last decade, in parallel to that of devices providing
mobile internet connections. Out of these online social networks, some of them
consider the data to be private while others emphasize in distributing the
information publicly. Out of the last ones, Twitter is, by far, the larger.
Tweets, intended as a compact way to share an opinion to the world, also contain
several kinds of information on the user and way it has been send. In particular,
it may include the geolocation at the time it is send. This information can be used
to study the mobility behaviour of people in urban areas, including attitudes and
lifestyle, which are particularly important for e.g. developing demand management
concepts for influencing mobility decisions.

The objective of this paper is to discuss efficient ways to retrieve and store
large amounts of data from social networks, such as Twitter. Since the fraction
of tweets that can be downloaded from Twitter is limited, besides a random sample,
we select the users which are located in the cities under study. We use Python
scripts to interact with Twitter API, in particular with Tweepy, and to decide
which are the users to be selected. To improve efficiency data is stored in a
distributed way using MongoDB noSQL data base. Finally we perform searches on the
database using the Python MongoDB driver to extract the relevant information to
study human mobility patterns. The paper is organized as follows, we first provide
a brief discussion on SQL and noSQL databases. The next section explains how data
is acquired and how users are selected. Then we compare the performance of a SQL
database with an distributed noSQL one. Finally, some preliminary mobility
results are shown and we give some concluding remarks.

\section{Data storage%
  \label{data-storage}%
}

Storage of data in plain files is not practical when dealing with thousands of
millions of tweets, even if each document is as small as a tweet. It is much more
convenient to store them in a database, which also allows for multi user-access
to the information. High availability, performance, scalability and recoverability,
together with good drivers, easy administration and preferably open source are
some of the desirable characteristics of a database to store large amounts of data
for a long period of time.

\subsection{SQL vs noSQL%
  \label{sql-vs-nosql}%
}

Traditional databases rely on collections of tables with data records, which
are formally described and organized according to a relational model. This is known
as Relational Database Management System, RMDBS) as introduced by E.F. Codd \cite{Codd}.
Relational databases are widely adopted and tested. As a consequence, Atomicity,
Consistency, Isolation and Durability (also known as ACID \cite{ACID}) is supported
as well as crash recovery, and master-slave consistency, which explains why they have been actively used in production since
the seventies. There are different tools to access relational databases but,
by far Structured Query Language (SQL) is the most common.
Therefore the term SQL is in many instances used as synonym to relational database.
On a relational database, data can not be inserted in an arbitrary way. One has
first to parse the data prior to insertion, which for fast data archival may
constitute a bottleneck. Typically a SQL database is stored in a pair of master-slave
servers. Since insertion and queries are performed on the same computer, performance
depends on the hardware of this computer and scalability involves improving
the existing hardware.

On the contrary, noSQL databases make use of less constrained consistency models
to gain simplicity of design, horizontal scalability and finer control over
availability. NoSQL databases are schema free and are capable of distributing
data among different computers. Insert and search throughput is related to the
number of computers in which the data is stored.
This gives them the ability to horizontally scale up, while relying on relatively
simple APIs. Therefore scalability is achieved by adding computers to the cluster
rather than improving the existing hardware.
Crash recovery and high availability are obtained by data replication, a configuration
similar to the well-known master-slave SQL configuration. Moreover, queries can be
distributed among the different computers that hold the data in order to
improve the performance.  All these make noSQL database management systems quite
useful when working with large
datasets or when the nature of the data does not require a relational model.

However, ACID which is guaranteed in SQL database transactions, is typically not
directly provided by noSQL databases. In particular, consistency is guaranteed
only after a period of time sufficient to propagate the changes through the
system. Thus, the management of noSQL databases is usually quite more cumbersome
than that of SQL ones.

Finally we would like to note than in order to improve the search performance on
both relational and non relational databases, it is important to index the data
fields which are used more frequently for classification.

\subsection{Documents format and storage%
  \label{documents-format-and-storage}%
}

Typically we download more than 15 million statuses (tweets) per day, and this
number grows over the time. However, the volume of tweets is not constant in time.
Political or social singular events can cause traffic spikes, thus the database must
have the ability to handle large and steadily increasing volume of data while
providing enough flexibility to deal with unexpected traffic peaks. In addition,
the format of the information contained in tweets can, and does, change over the time.

The data we retrieve from Twitter is either encoded as JavaScript Object Notation
(JSON), \cite{JSON} or it can be easily
converted to JSON format by using jsonpickle package \cite{jsonp}. JSON format
is used to transmit human readable data through a network connection as a
alternative to XML. The attributes of JSON encoded objects are unsorted and the
format of the attributes is not fixed. JSON flexibility allows for new fields to
be added or old fields to be
removed, and also to change the format of the fields (integer, string, datetime, ... ).

Scalability and read/write performance drive us to consider noSQL databases to store documents.
CouchDB \cite{CouchDB}, and MongoDB \cite{MongoDB}, have been considered in the study.
Both databases are quite similar. On both, replication and failover security is
achieved by replicating data on different servers. The advantages of
CouchDB include that it accepts JSON data, ACID performance and allowing the
use of map/reduce operations for query paralelization. However, queries have to
be predefined in advance and users can only perform predefined queries or
combinations of them. This is an inconvenient when the users of the database have
different needs and objectives. Besides the language for the queries is quite
specific and different from standard SQL commands.

On the other hand, MongoDB also accepts JSON data which is internally stored as
a BSON \cite{BSON} object, which is a JSON-like document in a binary format. MongoDB
does not require the queries to be predefined and furthermore it supports SQL-like
commands including aggregation of the results. Besides it also supports map/reduce
queries. Furthermore it provides a large variety of indexes including geo indices,
which is particular relevant for the kind of data queries we are interested in
since it allows queries to be executed in real time. Even that MongoDB lacks real
ACID transactions, it is more suitable when using large amounts of data. Horizontal
scalability is much clear in MongoDB and the possibility to allow users to run
their own queries without requiring predefinitions made MongoDB more suitable to our
needs. In what follows the only noSQL database that is considered is MongoDB.

\subsection{MongoDB configuration%
  \label{mongodb-configuration}%
}

MongoDB minimal configuration involves two computers, one server with all the data and
one client to which the user application connects to. This minimal configuration does not ensure failover recovery.
To ensure this, one needs additional computers forming a cluster. This cluster is
called Replica Set and these groups consist of a minimum of three computers one
of them designated as the primary and the others as secondaries. To ensure automatic
failover recovery, the members of the replica set run a daemon that replicates the data.
The primary member receives all the write/insert connections while secondaries replicate from the primary asynchronously
with a delay of milliseconds and can receive read orders. Even that data replication
uses much more space that the one really needed, it ensures high availability and increases read capacity. Apart from that,
a good practice is to configure one of the secondary computers of each replica
set with a predefined replication delay time and use it as backup. The standard
MongoDB configuration hides the backup computer from clients so that it can not
be used for searches to prevent searching in a non up to date data set.

The configuration with a single Replica Set is suitable when one single computer can store all
the data and the read/write performance using a single computer is enough.
In MongoDB the way to scale up the database is sharding: the collection of data is
partitioned by using a key defined by the database administrator
and the different chunks or portions are stored in different replica sets or shards.
Sharding automatically balances data and load across the shards and increases write
and read capacity. In addition to that, it provides a clear pathway to grow.
When a database collection becomes too large for the existing configuration, a
new shard (horizontal scalability) can be added and MongoDB
automatically distributes collection data to the new group of servers.\begin{figure}[]\noindent\makebox[\columnwidth][c]{\includegraphics[width=\columnwidth]{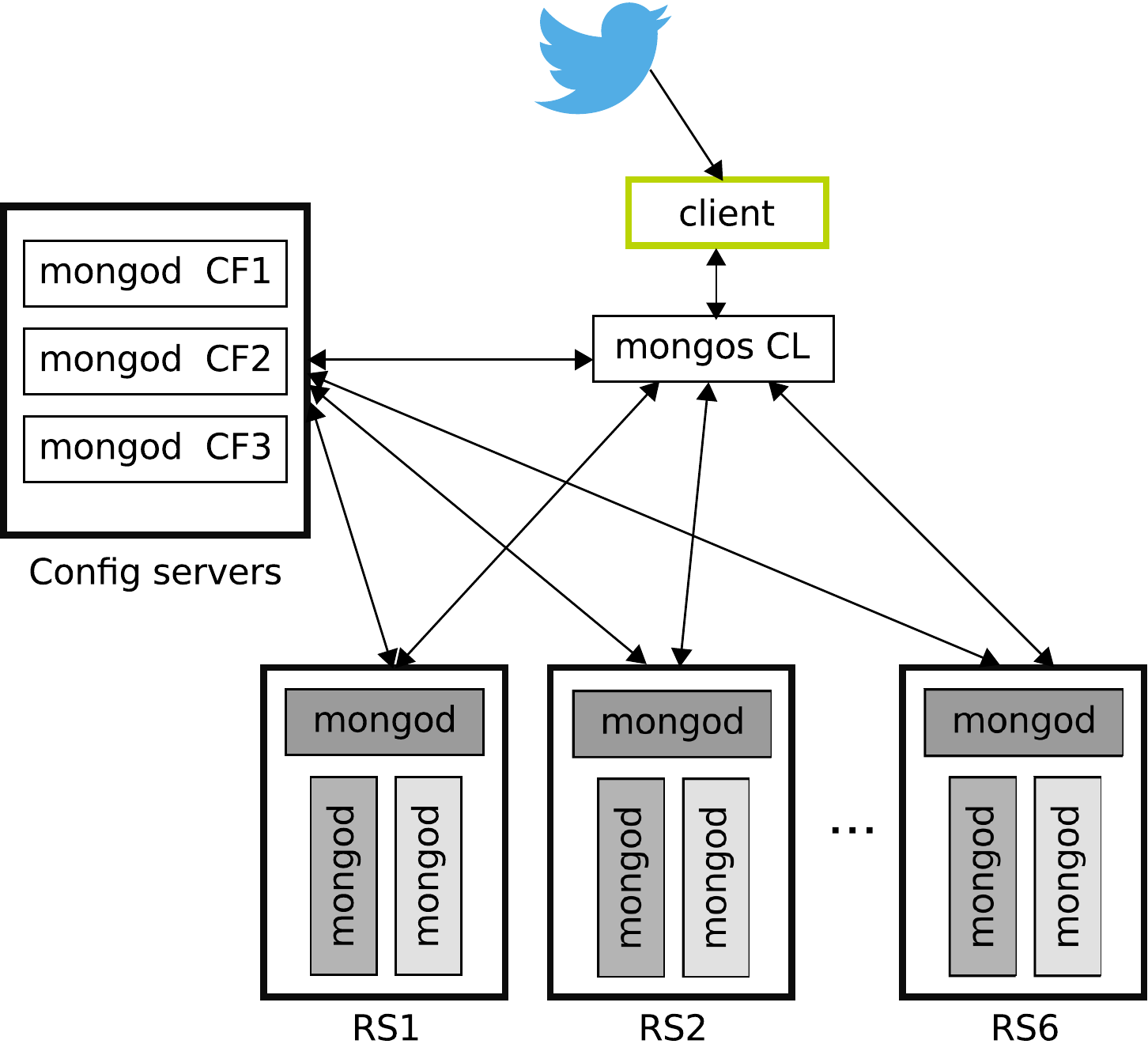}}
\caption{Schematic MongoDB configuration. \DUrole{label}{fig:config}}
\end{figure}

In addition to shards or replica sets, in a sharded cluster there are
configuration servers (CS) that store metadata relating replica sets with data portions and
that route reads and writes from mongo clients (CL) to the appropriate replica set. Notice
that client applications connect to mongo client (CL) which returns the answer
to the queries. The structure behind is hidden to the client.

In our case (Fig.  \DUrole{ref}{fig:config}), we configured a sharded cluster with
six replica sets formed by three members each. As recommended in production environments,
we are currently using three configuration servers.
Each of the replica sets has two eligible primary members and the third one
is a delayed copy by 72 hours. This gives us failover security because
if primary server daemon crashes or stops, the secondary one becomes primary.
The third member helps us to recover from human errors
such as inadvertently deleted databases or botched application upgrades.
Finally, the shard key used
is the tweet identifier and we added indices by user
identifier and latitude/longitude to speed up usual queries.

To improve writing performance we took into account several MongoDB
features when customizing the operating system in the servers that form
the replica sets.

\section{Data acquisition%
  \label{data-acquisition}%
}

Even that Twitter provides mechanisms to retrieve only a small fraction of total
amount of tweets (about 1\% randomly distributed), this constitutes a large amount
of data distributed all over the world. About 12\% of the retrieved tweets have
geolocated data. And only a small part of these are located in the cities we are
focusing in, such as London or Bacelona. As a consequence, the number of tweets
that can be used to study the human mobility in these cities is limited. To solve
this issue, we select a set of users from the random sample which have tweets
geolocated in the metropolitan areas and we download their timeline.

\subsection{Twitter APIs%
  \label{twitter-apis}%
}

Twitter data access can be achieved through two ad hoc APIs that represent
different Twitter features: Stream and Representational State Transfer
(REST). The '\emph{Stream API}' is focused in data mining providing the real time sample
of the tweets. The '\emph{REST API}' enables
developers to access some of the core primitives of Twitter including timelines,
status updates, and user information.

Although possible, directly access the Twitter APIs is not trivial. Therefore it
is recommended to use a library \cite{Twilib}. There are libraries for many computer
languages although here we focus on Python. The code readability, the smooth
learning curve, the quick development or the dynamic typing makes Python a suitable
language to be used by
software engineers and scientists. Among the libraries for Python we use the
\emph{tweepy} \cite{Tweepy}, for its simplicity and flexibility. Besides with a package
we can access both Stream and REST APIs. Furthermore, it is open source (MIT License).

The data we receive from streaming is JSON encoded while the data we gather
from other APIs is converted to JSON format by using jsonpickle package \cite{jsonp}.

\subsection{Random sample%
  \label{random-sample}%
}

Connecting to the streaming API requires having a Twitter account and keeping open
a persistent HTTP connection to one of the public endpoints. Streaming API do
not support requests. Randomly sampled tweets are provided automatically
(subject to the limitation of about 1\% of the total tweets). This API limits
each account to just one standing connection. In fact, connecting to a public
stream more than once with the same credentials causes the oldest connection to
be disconnected. Besides, IPs of clients that make excessive connection attempts
run the risk of being automatically banned.

As the process that opens the connection receives raw tweets, it has to perform
all parsing, filtering, and aggregation needed before storing the result.

A particularity of the \emph{Streaming API} is that messages are not
delivered in the same precise order as they were generated. In
particular, messages can be slightly shifted in time and it is also
possible that deleted messages are received before the original tweet.
This is not critical for the case considered here because we are
interested in slower time scales (from minutes to hours) and therefore
we do not need to have an exact timing and order of the messages.

In our case, connections to Twitter API are achieved by using \emph{tweepy}.
It allows the implementation of a listener that activates when a tweet arrives
and so that it can be processed. Prior to the connection to the API it is
necessary to provide the user
authentication and to instantiate the listener.

A example of the code to simply print the tweets to the standard output is:
\begin{DUlineblock}{0em}
\item[] 
\end{DUlineblock}
\begin{Verbatim}[commandchars=\\\{\},fontsize=\footnotesize]
\PY{k+kn}{from} \PY{n+nn}{tweepy} \PY{k+kn}{import} \PY{n}{Stream}\PY{p}{,} \PY{n}{OAuthHandler}
\PY{k+kn}{from} \PY{n+nn}{tweepy.streaming} \PY{k+kn}{import} \PY{n}{StreamListener}

\PY{k}{class} \PY{n+nc}{BasicListener}\PY{p}{(}\PY{n}{StreamListener}\PY{p}{)}\PY{p}{:}
  \PY{l+s+sd}{\PYZdq{}\PYZdq{}\PYZdq{}}
\PY{l+s+sd}{    A listener handles tweets are the received from}
\PY{l+s+sd}{    the stream.}
\PY{l+s+sd}{  \PYZdq{}\PYZdq{}\PYZdq{}}
  \PY{k}{def} \PY{n+nf}{on\PYZus{}data}\PY{p}{(}\PY{n+nb+bp}{self}\PY{p}{,} \PY{n}{data}\PY{p}{)}\PY{p}{:}
      \PY{c}{\PYZsh{} print received tweet to stdout}
      \PY{k}{print} \PY{n}{data}
      \PY{k}{return} \PY{n+nb+bp}{True}

  \PY{k}{def} \PY{n+nf}{on\PYZus{}error}\PY{p}{(}\PY{n+nb+bp}{self}\PY{p}{,} \PY{n}{status}\PY{p}{)}\PY{p}{:}
      \PY{c}{\PYZsh{} print error when data is not correctly}
      \PY{c}{\PYZsh{} received}
      \PY{k}{print} \PY{l+s}{\PYZdq{}}\PY{l+s}{Error: }\PY{l+s}{\PYZdq{}} \PY{o}{+} \PY{n}{status}

\PY{k}{if} \PY{n}{\PYZus{}\PYZus{}name\PYZus{}\PYZus{}} \PY{o}{==} \PY{l+s}{\PYZsq{}}\PY{l+s}{\PYZus{}\PYZus{}main\PYZus{}\PYZus{}}\PY{l+s}{\PYZsq{}}\PY{p}{:}

  \PY{c}{\PYZsh{} authentication}
  \PY{n}{auth} \PY{o}{=} \PY{n}{OAuthHandler}\PY{p}{(}\PY{n}{CONSUMER\PYZus{}KEY}\PY{p}{,} \PY{n}{CONSUMER\PYZus{}SECRET}\PY{p}{)}
  \PY{n}{auth}\PY{o}{.}\PY{n}{set\PYZus{}access\PYZus{}token}\PY{p}{(}\PY{n}{ACCESS\PYZus{}KEY}\PY{p}{,} \PY{n}{ACCESS\PYZus{}SECRET}\PY{p}{)}

  \PY{c}{\PYZsh{} listener instance}
  \PY{n}{listen} \PY{o}{=} \PY{n}{BasicListener}\PY{p}{(}\PY{p}{)}

  \PY{c}{\PYZsh{} open connection}
  \PY{n}{stream} \PY{o}{=} \PY{n}{Stream}\PY{p}{(}\PY{n}{auth}\PY{p}{,} \PY{n}{listen}\PY{p}{,} \PY{n}{gzip}\PY{o}{=}\PY{n+nb+bp}{True}\PY{p}{)}

  \PY{c}{\PYZsh{} start receiving data}
  \PY{n}{stream}\PY{o}{.}\PY{n}{sample}\PY{p}{(}\PY{p}{)}
\end{Verbatim}
In our case, we store the tweets we download in the same format they are received, JSON. To make data
search and manipulation easier we use the tweet id as one of the indices of the
database and it is necessary to have it in the highest level of the document.

Deleted tweets have a different structure than regular ones. In particular they are identified by the key, \emph{deleted},
in the highest document level. They only contain information on the user and the
tweetID of the erased tweet. We also store these tweets.

\subsection{Users selection%
  \label{users-selection}%
}

We identify the users which have tweets geolocated in the metropolitan areas
under study. Since we are interested in people that moves, we first disregard
all the users whose tweets come always from the same location. This also
filters out most of advertising tweets send by companies, since they have a
fixed
location. Then we sort users to be checked by number of geolocated tweets already
collected in order to prioritize the most active ones.

Identification of geotweets located in the areas of interest is achieved by using
\emph{geoNear} MongoDB command, \cite{geoNear} which returns the documents with a location
not exceeding a given distance (radius) from a given place. An example of how to
use geoNear command with MongoDB Python driver is%
\begin{quote}\begin{verbatim}
db.command(SON([('geoNear', collection),
                ('near', [lon, lat]),
                ('maxDistance', max_dist),
                ('num', max_num_results)
               ]))
\end{verbatim}

\end{quote}
MongoDB limits the size of the results document returned by the \emph{geoNear} query
to 16MB. To avoid exceeding this limitation we use a value for the radius of
exploration of one mile. To cover the metropolitan area we make use of a fine
grained mesh in which the points are separated by one mile and perform \emph{geoNear}
query at each grid point.

\subsection{Users wall%
  \label{users-wall}%
}

Retrieving the tweets posted by a specific user is done using user\_timelime method
of the \emph{REST API} to which we connect via tweepy. In order to connect to this API
we use an access token generated by Twitter. In current version of the API, 1.1,
there is a limit in the number of queries per access token and per method on a time
window  \cite{limit}. For the queries regarding user\_timelime the limit is set to 180
requests every fifteen minutes, \cite{timeline}

Some users have their tweets protected. This implies that while these tweets are
freely distributed via streaming, it is not possible to retrieve them via the
\emph{REST API}. Therefore when we detect that a user protects its tweets, we remove
him/her from the list of selected users.

The method \emph{user\_timeline} returns a collection of the most recent tweets posted
by the user indicated by the user\_id parameter. These tweets are stored in a
separate collection from the one used for the stream API on the same MongoDB
database. In order to retrieve the maximum possible tweets and to avoid having
duplicated data, we request tweets with an identifier higher than the highest
tweet id we have for that user. The \emph{user\_timeline} method can return a maximum
of 200 historical tweets per query and in the case of very active users some tweets can be lost.

An example of use on how to connect to the API and getting the timeline:\begin{Verbatim}[commandchars=\\\{\},fontsize=\footnotesize]
\PY{k+kn}{from} \PY{n+nn}{tweepy} \PY{k+kn}{import} \PY{n}{OAuthHandler}\PY{p}{,} \PY{n}{API}
\PY{o}{.}\PY{o}{.}\PY{o}{.}

\PY{n}{OAuth} \PY{o}{=} \PY{n}{OAuthHandler}\PY{p}{(}\PY{n}{CONSUMER\PYZus{}KEY}\PY{p}{,} \PY{n}{CONSUMER\PYZus{}SECRET}\PY{p}{)}
\PY{n}{OAuth}\PY{o}{.}\PY{n}{set\PYZus{}access\PYZus{}token}\PY{p}{(}\PY{n}{ACCESS\PYZus{}KEY}\PY{p}{,} \PY{n}{ACCESS\PYZus{}SECRET}\PY{p}{)}
\PY{n}{tAPI} \PY{o}{=} \PY{n}{API}\PY{p}{(}\PY{n}{OAuth}\PY{p}{)}

\PY{n}{timeline} \PY{o}{=} \PY{n}{tAPI}\PY{o}{.}\PY{n}{user\PYZus{}timeline}\PY{p}{(}\PY{n}{count}\PY{o}{=}\PY{l+m+mi}{200}\PY{p}{,} \PY{n}{user\PYZus{}id}\PY{o}{=}\PY{n}{uid}\PY{p}{,}
                              \PY{n}{since\PYZus{}id}\PY{o}{=}\PY{n}{last\PYZus{}id}\PY{p}{)}
\end{Verbatim}

\subsection{Users network%
  \label{users-network}%
}
Retrieving the user network can be also done through the \emph{REST API}.
To do so we query for the list of user identifiers the specified user is following,
or \cite{friends}, and for the list of users the user follows
(reads the wall,  \cite{follow}) at the moment we do the query.

Queries to each method are limited to 15 requests every fifteen minutes. In order
to study the network evolution, we perform periodic queries for the different
users within the limitations the API imposes.

The code to get the followers and friends is:\begin{Verbatim}[commandchars=\\\{\},fontsize=\footnotesize]
\PY{n}{tAPI}\PY{o}{.}\PY{n}{friends\PYZus{}ids}\PY{p}{(}\PY{n}{uid}\PY{p}{)}
\PY{n}{tAPI}\PY{o}{.}\PY{n}{followers\PYZus{}ids}\PY{p}{(}\PY{n}{uid}\PY{p}{)}
\end{Verbatim}

\section{Data insertion%
  \label{data-insertion}%
}
We discuss here the procedure to insert the JSON documents retrieved from Twitter
into a relational database, such as MySQL, and in MongoDB.

\subsection{MySQL%
  \label{mysql}%
}

When using MySQL, JSON data cannot be directly inserted into the database.
Therefore it is necessary to parse JSON to a relational model. In particular we have to map the fields in the
JSON document to variables in classes.
To do that, we take advantage of Django Object Relational Model, ORM, \cite{Django}.
First, one creates an empty Django project and then, one sets the database connection information
in the project configuration file, settings.py. The database information to be included
in settings.py is the following:\begin{Verbatim}[commandchars=\\\{\},fontsize=\footnotesize]
\PY{n}{DATABASES} \PY{o}{=} \PY{p}{\PYZob{}}
    \PY{l+s}{\PYZsq{}}\PY{l+s}{default}\PY{l+s}{\PYZsq{}}\PY{p}{:} \PY{p}{\PYZob{}}
        \PY{l+s}{\PYZsq{}}\PY{l+s}{ENGINE}\PY{l+s}{\PYZsq{}}\PY{p}{:} \PY{l+s}{\PYZsq{}}\PY{l+s}{django.db.backends.mysql}\PY{l+s}{\PYZsq{}}\PY{p}{,}
        \PY{l+s}{\PYZsq{}}\PY{l+s}{NAME}\PY{l+s}{\PYZsq{}}\PY{p}{:} \PY{l+s}{\PYZsq{}}\PY{l+s}{twitterdb}\PY{l+s}{\PYZsq{}}\PY{p}{,}
        \PY{l+s}{\PYZsq{}}\PY{l+s}{USER}\PY{l+s}{\PYZsq{}}\PY{p}{:} \PY{l+s}{\PYZsq{}}\PY{l+s}{theuser}\PY{l+s}{\PYZsq{}}\PY{p}{,}
        \PY{l+s}{\PYZsq{}}\PY{l+s}{PASSWORD}\PY{l+s}{\PYZsq{}}\PY{p}{:} \PY{l+s}{\PYZsq{}}\PY{l+s}{thepassword}\PY{l+s}{\PYZsq{}}\PY{p}{,}
        \PY{l+s}{\PYZsq{}}\PY{l+s}{HOST}\PY{l+s}{\PYZsq{}}\PY{p}{:} \PY{l+s}{\PYZsq{}}\PY{l+s}{mysqlHost}\PY{l+s}{\PYZsq{}}\PY{p}{,}
        \PY{l+s}{\PYZsq{}}\PY{l+s}{PORT}\PY{l+s}{\PYZsq{}}\PY{p}{:} \PY{l+s}{\PYZsq{}}\PY{l+s}{3360}\PY{l+s}{\PYZsq{}}\PY{p}{,}
    \PY{p}{\PYZcb{}}
\PY{p}{\PYZcb{}}
\end{Verbatim}
In the project's application, one creates a relational model with some classes
(Tweet, User, HashTag, URL, ... ). Primary keys and relations between registers
are used to avoid data duplication. An example of the Tweet model class, which is the main one, is:\begin{Verbatim}[commandchars=\\\{\},fontsize=\footnotesize]
\PY{k}{class} \PY{n+nc}{Tweet}\PY{p}{(}\PY{n}{Model}\PY{p}{)}\PY{p}{:}
    \PY{n}{twid} \PY{o}{=} \PY{n}{BigIntegerField}\PY{p}{(}\PY{n}{primary\PYZus{}key}\PY{o}{=}\PY{n+nb+bp}{True}\PY{p}{,}
                           \PY{n}{db\PYZus{}index}\PY{o}{=}\PY{n+nb+bp}{True}\PY{p}{)}
    \PY{n}{place} \PY{o}{=} \PY{n}{ForeignKey}\PY{p}{(}\PY{n}{Place}\PY{p}{,} \PY{n}{null}\PY{o}{=}\PY{n+nb+bp}{True}\PY{p}{)}
    \PY{n}{text} \PY{o}{=} \PY{n}{CharField}\PY{p}{(}\PY{n}{max\PYZus{}length}\PY{o}{=}\PY{l+m+mi}{2048}\PY{p}{,} \PY{n}{blank}\PY{o}{=}\PY{n+nb+bp}{True}\PY{p}{)}
    \PY{n}{retweet\PYZus{}count} \PY{o}{=} \PY{n}{IntegerField}\PY{p}{(}\PY{n}{null}\PY{o}{=}\PY{n+nb+bp}{True}\PY{p}{)}
    \PY{n}{parent\PYZus{}id} \PY{o}{=} \PY{n}{BigIntegerField}\PY{p}{(}\PY{n}{null}\PY{o}{=}\PY{n+nb+bp}{True}\PY{p}{)}
    \PY{n}{source} \PY{o}{=} \PY{n}{CharField}\PY{p}{(}\PY{n}{max\PYZus{}length}\PY{o}{=}\PY{l+m+mi}{2048}\PY{p}{)}
    \PY{n}{coordinates} \PY{o}{=} \PY{n}{ForeignKey}\PY{p}{(}\PY{n}{BoundingBox}\PY{p}{,} \PY{n}{null}\PY{o}{=}\PY{n+nb+bp}{True}\PY{p}{)}
    \PY{n}{contributors} \PY{o}{=} \PY{n}{CharField}\PY{p}{(}\PY{n}{max\PYZus{}length}\PY{o}{=}\PY{l+m+mi}{2048}\PY{p}{,}
                             \PY{n}{null}\PY{o}{=}\PY{n+nb+bp}{True}\PY{p}{)}
    \PY{n}{retweeted} \PY{o}{=} \PY{n}{BooleanField}\PY{p}{(}\PY{p}{)}
    \PY{n}{truncated} \PY{o}{=} \PY{n}{BooleanField}\PY{p}{(}\PY{p}{)}
    \PY{n}{created\PYZus{}at} \PY{o}{=} \PY{n}{DateTimeField}\PY{p}{(}\PY{n}{null}\PY{o}{=}\PY{n+nb+bp}{True}\PY{p}{)}
    \PY{n}{user} \PY{o}{=} \PY{n}{ForeignKey}\PY{p}{(}\PY{n}{User}\PY{p}{)}
    \PY{n}{entities} \PY{o}{=} \PY{n}{ForeignKey}\PY{p}{(}\PY{n}{Entities}\PY{p}{,} \PY{n}{null}\PY{o}{=}\PY{n+nb+bp}{True}\PY{p}{)}
    \PY{n}{in\PYZus{}reply\PYZus{}to\PYZus{}status\PYZus{}id} \PY{o}{=} \PY{n}{BigIntegerField}\PY{p}{(}
                              \PY{n}{null}\PY{o}{=}\PY{n+nb+bp}{True}\PY{p}{)}
    \PY{n}{in\PYZus{}reply\PYZus{}to\PYZus{}user\PYZus{}id} \PY{o}{=} \PY{n}{BigIntegerField}\PY{p}{(}
                              \PY{n}{null}\PY{o}{=}\PY{n+nb+bp}{True}\PY{p}{)}
    \PY{n}{in\PYZus{}reply\PYZus{}to\PYZus{}screen\PYZus{}id} \PY{o}{=} \PY{n}{BigIntegerField}\PY{p}{(}
                              \PY{n}{null}\PY{o}{=}\PY{n+nb+bp}{True}\PY{p}{)}
    \PY{n}{deleted} \PY{o}{=} \PY{n}{BooleanField}\PY{p}{(}\PY{p}{)}

    \PY{k}{class} \PY{n+nc}{Meta}\PY{p}{:}
        \PY{n}{app\PYZus{}label} \PY{o}{=} \PY{l+s}{\PYZsq{}}\PY{l+s}{twitter}\PY{l+s}{\PYZsq{}}
\end{Verbatim}
Similarly for the class User, HashTag, URL, ...

Note that when generating the user key the Tweet class we make use of the ForeignKey
class, so that if the user has already given a key, no new key is generated,
instead a link to the existing register is performed. While this reduces the
volume of data to be stored, it implies a search for each new tweet on the user
index of the database to find out if the user is already there. Similarly we also
use ForeignKey for place, coordinates and entities keys and therefore it is
necessary to do a search on the corresponding index of the database for each new
tweet.

Finally, for every JSON document, a parsing function is needed to store the data into
the database. A sample of the parsing function is:\begin{Verbatim}[commandchars=\\\{\},fontsize=\footnotesize]
\PY{k}{def} \PY{n+nf}{fillTweet}\PY{p}{(}\PY{n}{jsondata}\PY{p}{)}\PY{p}{:}
  \PY{n}{t} \PY{o}{=} \PY{n}{Tweet}\PY{p}{(}\PY{p}{)}

  \PY{k}{if} \PY{n}{propertyExists}\PY{p}{(}\PY{n}{jsondata}\PY{p}{,} \PY{l+s}{\PYZdq{}}\PY{l+s}{delete}\PY{l+s}{\PYZdq{}}\PY{p}{)}\PY{p}{:}
    \PY{n}{logger}\PY{o}{.}\PY{n}{info}\PY{p}{(}\PY{l+s}{\PYZdq{}}\PY{l+s}{Deleted tweet}\PY{l+s}{\PYZdq{}}\PY{p}{)}
    \PY{c}{\PYZsh{} do some magic}
  \PY{k}{else}\PY{p}{:}
    \PY{n}{logger}\PY{o}{.}\PY{n}{info}\PY{p}{(}\PY{l+s}{\PYZdq{}}\PY{l+s}{New tweet}\PY{l+s}{\PYZdq{}}\PY{p}{)}

    \PY{n}{twlist} \PY{o}{=} \PY{n}{Tweet}\PY{o}{.}\PY{n}{objects}\PY{o}{.}\PY{n}{filter}\PY{p}{(}
                          \PY{n}{twid}\PY{o}{=}\PY{n}{jsondata}\PY{p}{[}\PY{l+s}{\PYZdq{}}\PY{l+s}{id}\PY{l+s}{\PYZdq{}}\PY{p}{]}\PY{p}{)}
    \PY{k}{if} \PY{n+nb}{len}\PY{p}{(}\PY{n}{twlist}\PY{p}{)} \PY{o}{==} \PY{l+m+mi}{1}\PY{p}{:}
      \PY{n}{logger}\PY{o}{.}\PY{n}{info}\PY{p}{(}\PY{l+s}{\PYZdq{}}\PY{l+s}{already added}\PY{l+s}{\PYZdq{}}\PY{p}{)}
      \PY{k}{return} \PY{n}{twlist}\PY{p}{[}\PY{l+m+mi}{0}\PY{p}{]}\PY{o}{.}\PY{n}{twid}

    \PY{n}{t}\PY{o}{.}\PY{n}{contributors} \PY{o}{=} \PY{n}{fillContributors}\PY{p}{(}\PY{n}{jsondata}\PY{p}{)}
    \PY{n}{t}\PY{o}{.}\PY{n}{coordinatespt} \PY{o}{=} \PY{n}{fillPointBBox}\PY{p}{(}\PY{n}{jsondata}\PY{p}{)}
    \PY{n}{t}\PY{o}{.}\PY{n}{created\PYZus{}at} \PY{o}{=} \PY{n}{fillCreatedAt}\PY{p}{(}\PY{n}{jsondata}\PY{p}{)}
    \PY{n}{t}\PY{o}{.}\PY{n}{entities} \PY{o}{=} \PY{n}{fillEntities}\PY{p}{(}\PY{n}{jsondata}\PY{p}{)}
    \PY{n}{t}\PY{o}{.}\PY{n}{in\PYZus{}reply\PYZus{}to\PYZus{}screen\PYZus{}id} \PY{o}{=}
                \PY{n}{fillReplyScreen}\PY{p}{(}\PY{n}{jsondata}\PY{p}{)}
    \PY{n}{t}\PY{o}{.}\PY{n}{in\PYZus{}reply\PYZus{}to\PYZus{}status\PYZus{}id} \PY{o}{=}
                \PY{n}{fillReplyStatus}\PY{p}{(}\PY{n}{jsondata}\PY{p}{)}
    \PY{n}{t}\PY{o}{.}\PY{n}{in\PYZus{}reply\PYZus{}to\PYZus{}user\PYZus{}id} \PY{o}{=}
                \PY{n}{fillReplyUser}\PY{p}{(}\PY{n}{jsondata}\PY{p}{)}
    \PY{n}{t}\PY{o}{.}\PY{n}{place} \PY{o}{=} \PY{n}{fillPlace}\PY{p}{(}\PY{n}{jsondata}\PY{p}{)}
    \PY{n}{t}\PY{o}{.}\PY{n}{retweet\PYZus{}count} \PY{o}{=} \PY{n}{fillRTCount}\PY{p}{(}\PY{n}{jsondata}\PY{p}{)}
    \PY{n}{t}\PY{o}{.}\PY{n}{retweeted} \PY{o}{=} \PY{n}{fillRT}\PY{p}{(}\PY{n}{jsondta}\PY{p}{)}
    \PY{n}{t}\PY{o}{.}\PY{n}{source} \PY{o}{=} \PY{n}{fillSource}\PY{p}{(}\PY{n}{jsondata}\PY{p}{)}
    \PY{n}{t}\PY{o}{.}\PY{n}{text} \PY{o}{=} \PY{n}{fillText}\PY{p}{(}\PY{n}{jsondata}\PY{p}{)}
    \PY{n}{t}\PY{o}{.}\PY{n}{truncated} \PY{o}{=} \PY{n}{fillTruncated}\PY{p}{(}\PY{n}{jsondata}\PY{p}{)}
    \PY{n}{t}\PY{o}{.}\PY{n}{twid} \PY{o}{=} \PY{n}{jsondata}\PY{p}{[}\PY{l+s}{\PYZdq{}}\PY{l+s}{id}\PY{l+s}{\PYZdq{}}\PY{p}{]}
    \PY{n}{t}\PY{o}{.}\PY{n}{user} \PY{o}{=} \PY{n}{fillUser}\PY{p}{(}\PY{n}{jsondata}\PY{p}{)}

    \PY{c}{\PYZsh{} is retweet}
    \PY{n}{rtstatus} \PY{o}{=} \PY{l+s}{\PYZdq{}}\PY{l+s}{retweeted\PYZus{}status}\PY{l+s}{\PYZdq{}}
    \PY{k}{if} \PY{n}{propertyExists}\PY{p}{(}\PY{n}{jsondata}\PY{p}{,} \PY{n}{rtstatus}\PY{p}{)}\PY{p}{:}
      \PY{n}{logger}\PY{o}{.}\PY{n}{info}\PY{p}{(}\PY{l+s}{\PYZdq{}}\PY{l+s}{Is Retweet of }\PY{l+s}{\PYZdq{}}\PY{p}{)}
      \PY{n}{rtdata} \PY{o}{=} \PY{n}{jsondata}\PY{p}{[}\PY{n}{rtstatus}\PY{p}{]}
      \PY{n}{t}\PY{o}{.}\PY{n}{parent\PYZus{}id} \PY{o}{=} \PY{n}{fillTweet}\PY{p}{(}\PY{n}{rtdata}\PY{p}{)}

    \PY{c}{\PYZsh{} use Django DB connection to save to DB}
    \PY{n}{t}\PY{o}{.}\PY{n}{save}\PY{p}{(}\PY{p}{)}
\end{Verbatim}

\subsection{MongoDB%
  \label{mongodb}%
}
Connections to MongoDB can be done using \emph{pymongo}, \cite{pymongo}, the official driver for Python:\begin{Verbatim}[commandchars=\\\{\},fontsize=\footnotesize]
\PY{n}{mongoserver\PYZus{}uri} \PY{o}{=} \PY{l+s}{\PYZdq{}}\PY{l+s}{mongodb://}\PY{l+s}{\PYZdq{}} \PY{o}{+} \PY{n}{user} \PY{o}{+} \PY{l+s}{\PYZdq{}}\PY{l+s}{:}\PY{l+s}{\PYZdq{}} \PY{o}{+}
                  \PY{n}{pwd} \PY{o}{+} \PY{l+s}{\PYZdq{}}\PY{l+s}{@}\PY{l+s}{\PYZdq{}} \PY{o}{+} \PY{n}{host} \PY{o}{+} \PY{l+s}{\PYZdq{}}\PY{l+s}{:}\PY{l+s}{\PYZdq{}} \PY{o}{+}
                  \PY{n}{port} \PY{o}{+} \PY{l+s}{\PYZdq{}}\PY{l+s}{/}\PY{l+s}{\PYZdq{}} \PY{o}{+} \PY{n}{dbname}
\PY{n}{conection} \PY{o}{=} \PY{n}{MongoClient}\PY{p}{(}\PY{n}{host}\PY{o}{=}\PY{n}{mongoserver\PYZus{}uri}\PY{p}{)}
\PY{n}{db} \PY{o}{=} \PY{n}{conection}\PY{p}{[}\PY{n}{dbname}\PY{p}{]}
\PY{n}{collection} \PY{o}{=} \PY{n}{db}\PY{p}{[}\PY{n}{collectionname}\PY{p}{]}
\end{Verbatim}
Where \emph{collectionname} designs the collection to which the documents are to be inserted.

Since MongoDB accepts JSON documents no preprocessing is needed for the tweets
downloaded from the Streaming API. So, one connects to the stream collection and inserts directly the document:\begin{Verbatim}[commandchars=\\\{\},fontsize=\footnotesize]
\PY{n}{collection}\PY{o}{.}\PY{n}{insert}\PY{p}{(}\PY{n}{json\PYZus{}tweet}\PY{p}{)}
\end{Verbatim}
The user\_timeline method of the REST API returns a Python object which can be easily
converted to a JSON document and then inserted in the timeline collection.\begin{Verbatim}[commandchars=\\\{\},fontsize=\footnotesize]
\PY{n}{pickled} \PY{o}{=} \PY{n}{jsonpickle}\PY{o}{.}\PY{n}{encode}\PY{p}{(}\PY{n}{python\PYZus{}tweet}\PY{p}{)}
\PY{n}{json\PYZus{}tweet} \PY{o}{=} \PY{n}{json}\PY{o}{.}\PY{n}{loads}\PY{p}{(}\PY{n}{pickled}\PY{p}{)}
\PY{n}{collection}\PY{o}{.}\PY{n}{insert}\PY{p}{(}\PY{n}{json\PYZus{}tweet}\PY{p}{)}
\end{Verbatim}

\section{Database insertion performance%
  \label{database-insertion-performance}%
}
We first analyse the insertion rate performance for MySQL and MongoDB. The physical
computers we used had two Xeon L5520 at 2.27GHz processors with a total of 8 cores, 16GB of DDR3 RAM and a 2TB
hard disk (7200rpm).

\subsection{MySQL%
  \label{id18}%
}
\begin{figure}[]\noindent\makebox[\columnwidth][c]{\includegraphics[width=\columnwidth]{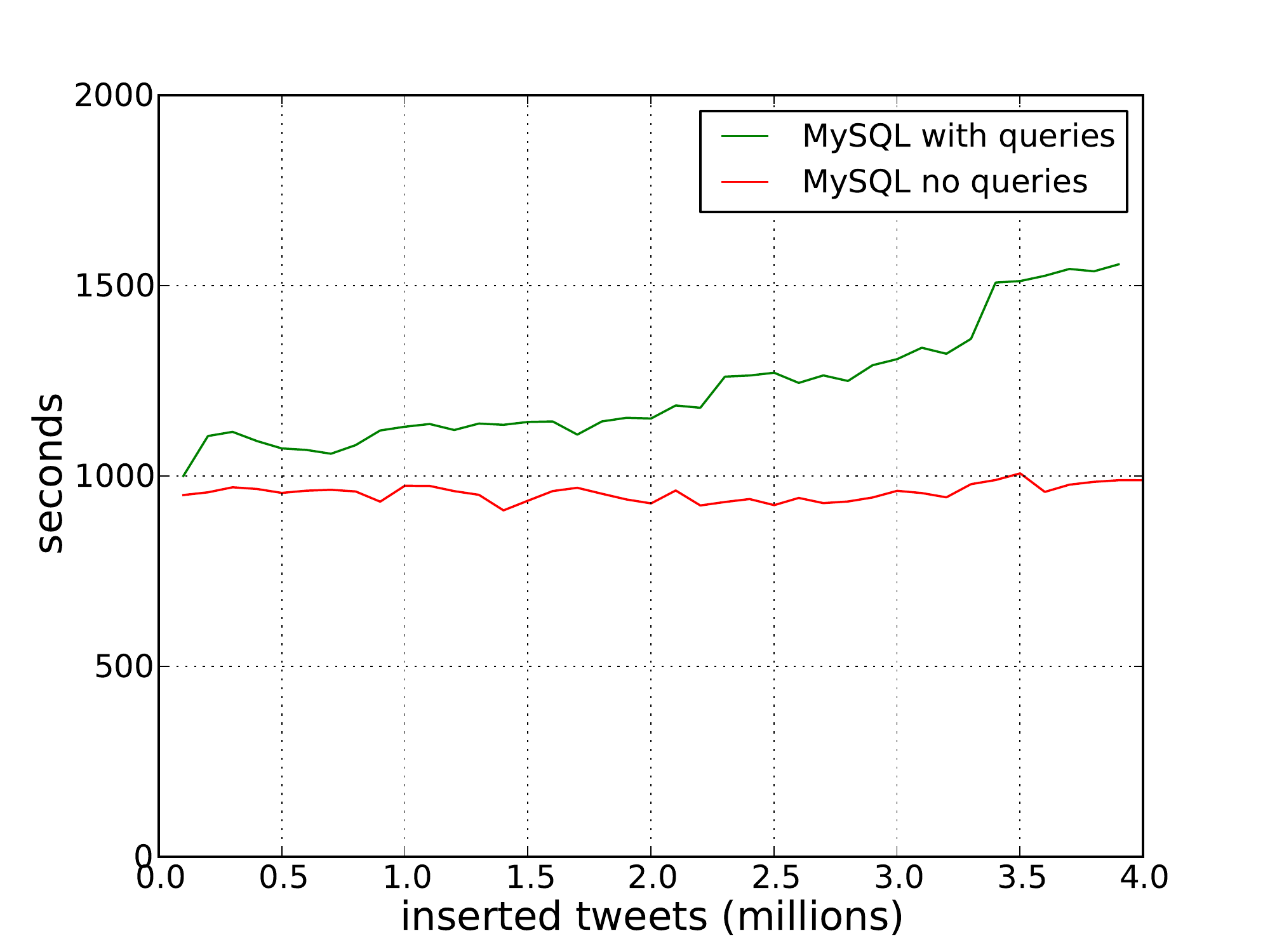}}
\caption{Time to insert 100000 tweets in MySQL using an empty database and tweets
processed with ORM. Linking (green) and duplicating information (red).
\DUrole{label}{fig:mysqlinsertion}}
\end{figure}

In Figure \DUrole{ref}{fig:mysqlinsertion} the green line shows the time to insert 100000
tweets in a completely empty MySQL database running on a single
physical computer. As explained above, when inserting tweets in MySQL,
as it is a relational database, we first perform several searches to find if the Twitter low level entities
such as user, hashtag, URL, ... exist, which results in a larger storage time.
As shown in \DUrole{ref}{fig:mysqlinsertion}, as the database grows the searches take longer and the
insertion rate decreases. It takes 1000 s when it is empty, above 1500 s
when there are four million tweets and almost 4000 s when the database
has twelve million tweets.

To avoid this search issue, we tested the same insertion procedure without using
ForeignKey, so that no searches are performed. For instance the corresponding lines of the Tweet class would read:\begin{Verbatim}[commandchars=\\\{\},fontsize=\footnotesize]
\PY{n}{coordinates} \PY{o}{=} \PY{n}{BoundingBox}\PY{p}{(}\PY{p}{)}
\PY{o}{.}\PY{o}{.}\PY{o}{.}
\PY{n}{user} \PY{o}{=} \PY{n}{User}\PY{p}{(}\PY{p}{)}
\PY{n}{entities} \PY{o}{=} \PY{n}{Entities}\PY{p}{(}\PY{p}{)}
\end{Verbatim}
As a consequence, in this case one does not takes advantage of the relational
properties of the database and duplicates data. The results are shown as a red
line in Fig. \DUrole{ref}{fig:mysqlinsertion}. The insertion rate is almost constant while
inserting 4 million tweets. The insertion rate, in fact, slightly reduces when the
database is over 15 million tweets.

\subsection{MongoDB%
  \label{id19}%
}
\begin{figure}[]\noindent\makebox[\columnwidth][c]{\includegraphics[width=\columnwidth]{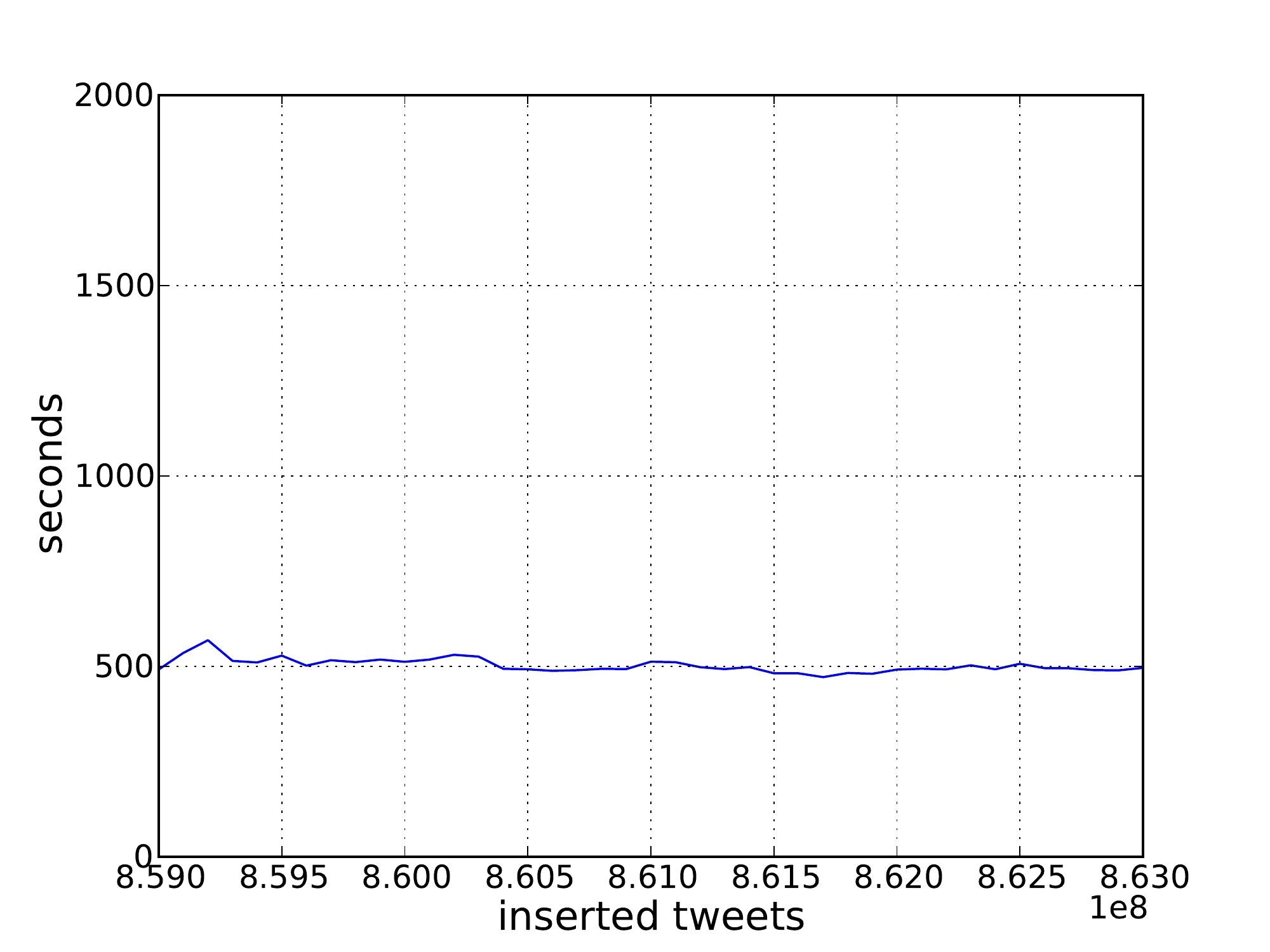}}
\caption{Time to insert 100000 tweets in MongoDB by using direct insertion in a database
with millions of tweets. \DUrole{label}{fig:mongoinsert}}
\end{figure}

Figure \DUrole{ref}{fig:mongoinsert} shows the time needed to insert 100000 tweets in a MongoDB
database with three shards (replica sets) using a single client. Here, instead
of starting with an empty database, we tested the performance with a database that had already 850 million documents stored.
As can be seen, storage time is much smaller that in MySQL, around 500 s for the
100000 tweets, which
is a speed up factor of two with respect to the MySQL when no search is
performed. Although the speed up is smaller than the factor three
expected from the fact of having three replica sets, it is still
substantial. What is more important, since we do not need to perform
searches, this performance is maintained as the database size grows.\begin{figure}[]\noindent\makebox[\columnwidth][c]{\includegraphics[width=\columnwidth]{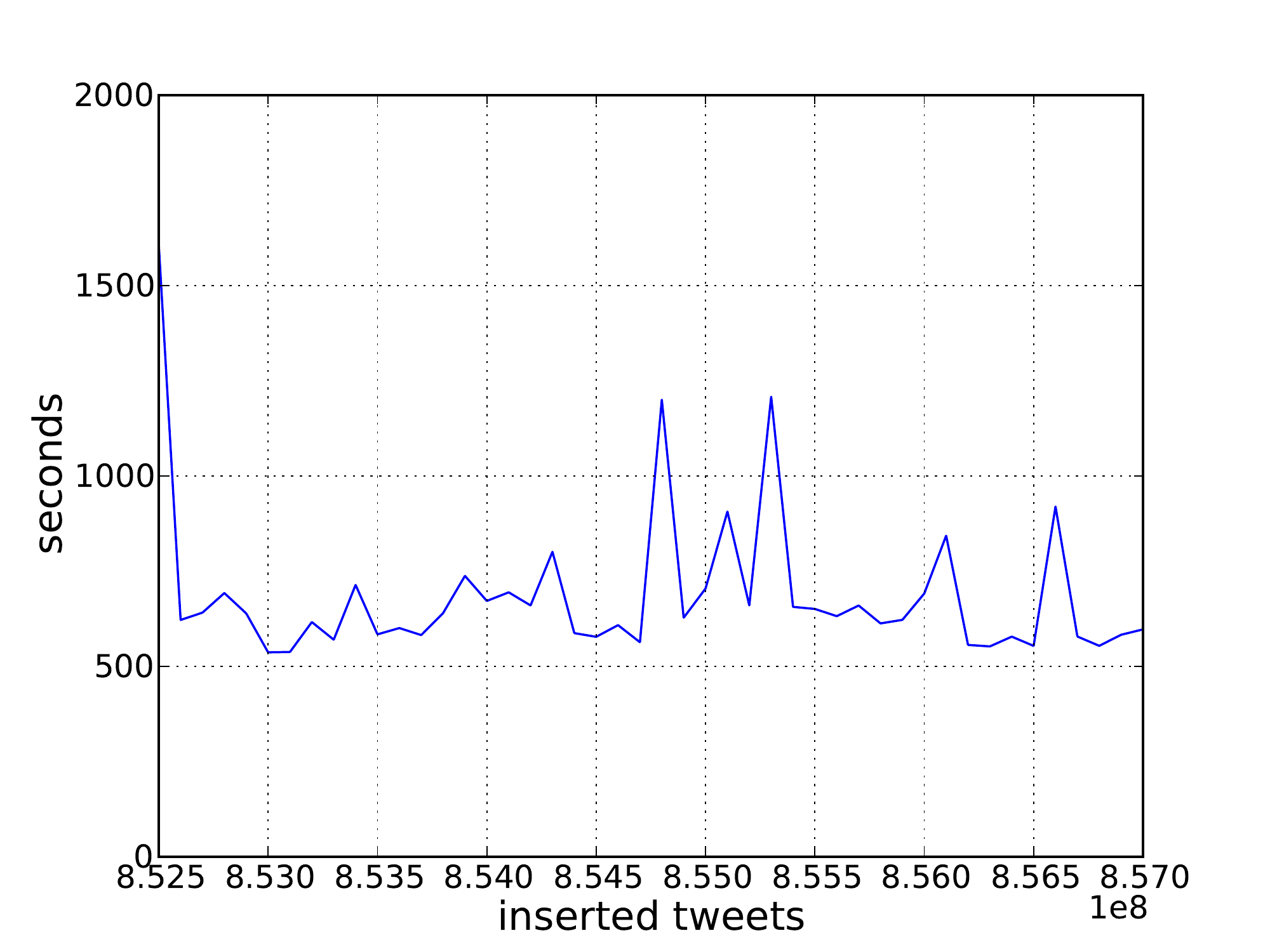}}
\caption{MongoDB insertion with a database with millions of tweets while querying the DB with CPU and memory consuming geoqueries. \DUrole{label}{fig:queriesinsertion}}
\end{figure}

\section{MongoDB query performance%
  \label{mongodb-query-performance}%
}

Here we analyse the response time of MongoDB when queries are performed.
MongoDB allows to configure the queries to be performed on the primary nodes or on the secondary ones.
This flexibility is particularly convenient in situations where data is continuously inserted, as the case considered here,
since one can configure the system so that queries are performed on the secondary nodes leaving the insertion data rate unaffected.

Since our goal is to analyse the geolocated tweets stored in the database we focus on MongoDB spatial indexing and querying.
MongoDB offers a specific geospatial index 2d for data stored as points on a
two-dimensional plane. The 2d index supports distance calculations on a sphere but not more complex calculations.
As of version 2.4 MongoDB also includes the index 2dsphere which conveniently
supports queries that calculate geometries on a sphere. This index supports data
stored as GeoJSON objects which is the way geospatial data is stored in the tweets.
Even that 2dsphere index seems to be more suitable that 2d one, in version 2.4
2dsphere indexing does not allow sparse data.
The GeoJSON object needs to exist in all the documents, but as as discussed before,
only a fraction of the tweets are geolocated. This bug has been solved and the
solution will be implemented in the upcoming 2.6 version.
Until then, we are using 2d indices (latitude, longitude) to determine the localization of the user when the tweet was posted.\begin{figure}[]\noindent\makebox[\columnwidth][c]{\includegraphics[width=\columnwidth]{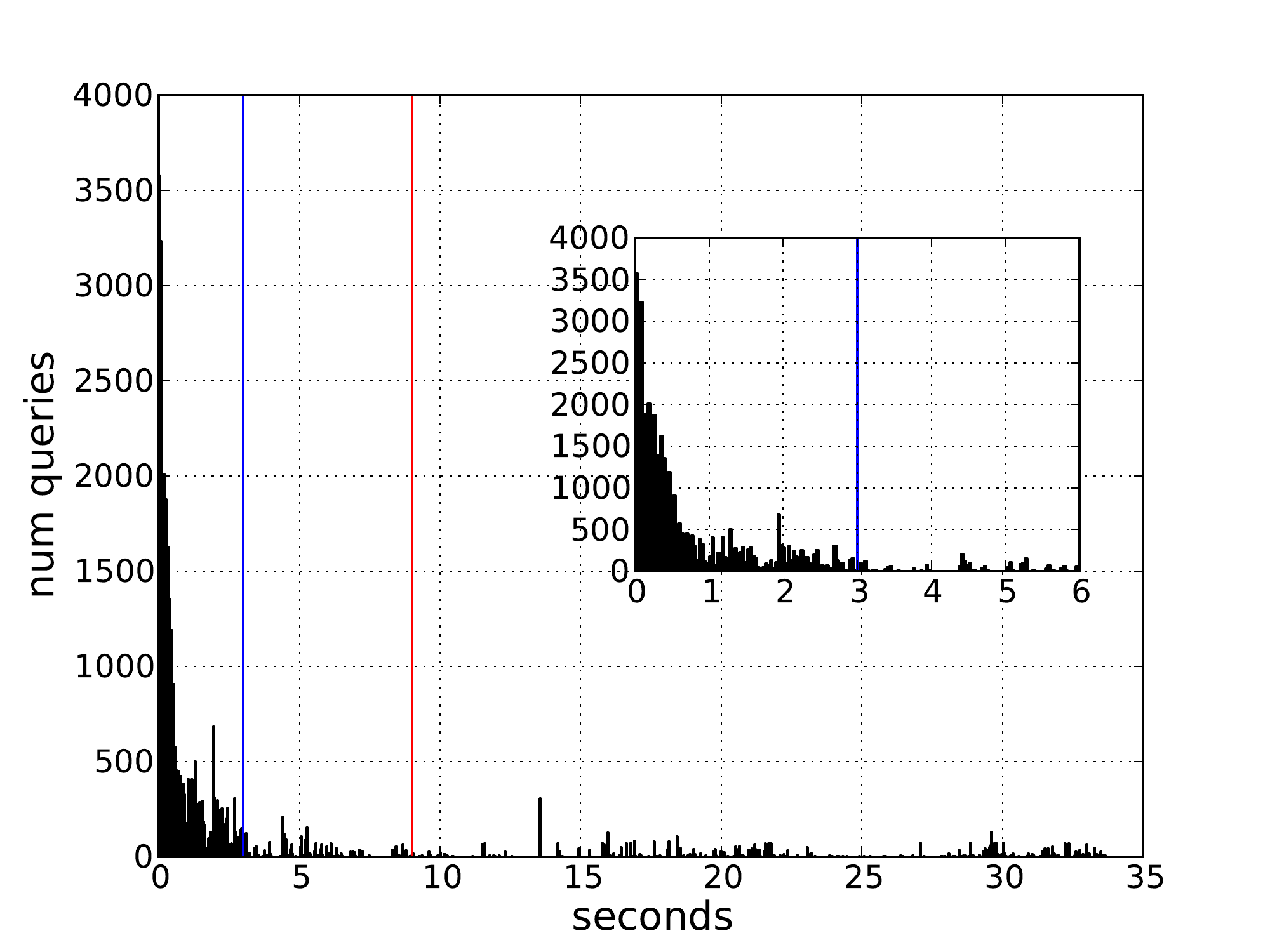}}
\caption{Queries timing histogram for Barcelona metropolitan area. Blue line shows the median and red line the 70th percentile. \DUrole{label}{fig:queries}}
\end{figure}

As discussed above we consider a mesh of points separated by one mile to cover
the metropolitan area under study and perform a \emph{geoNear} query at each grid point.
Fig. \DUrole{ref}{fig:queries} shows the histogram of the response time to \emph{geoNear}
queries in a database with 850 million documents stored. Although the response
time to a small group queries was slow response, more than thirty seconds, the
average (red blue line) is just of three seconds, and in 70\% of the queries to
get the tweets localized in a radius of one mile of a given point lasted less
than nine seconds (red vertical line).

\section{Preliminary results for mobility patterns%
  \label{preliminary-results-for-mobility-patterns}%
}

Preliminary results after retrieving data for ten months show a good
agreement with population in London and Barcelona metropolitan areas and the
transportation network of these cities (see Fig. \DUrole{ref}{fig:bcn} and \DUrole{ref}{fig:lon}).
This means that ten months sampling is representative of the mobility in these
areas.

In order to further assess that the data is statistically correct we plan to compare the statistics obtained from
the first ten months with the ones obtained after two years.\begin{figure}[]\noindent\makebox[\columnwidth][c]{\includegraphics[width=\columnwidth]{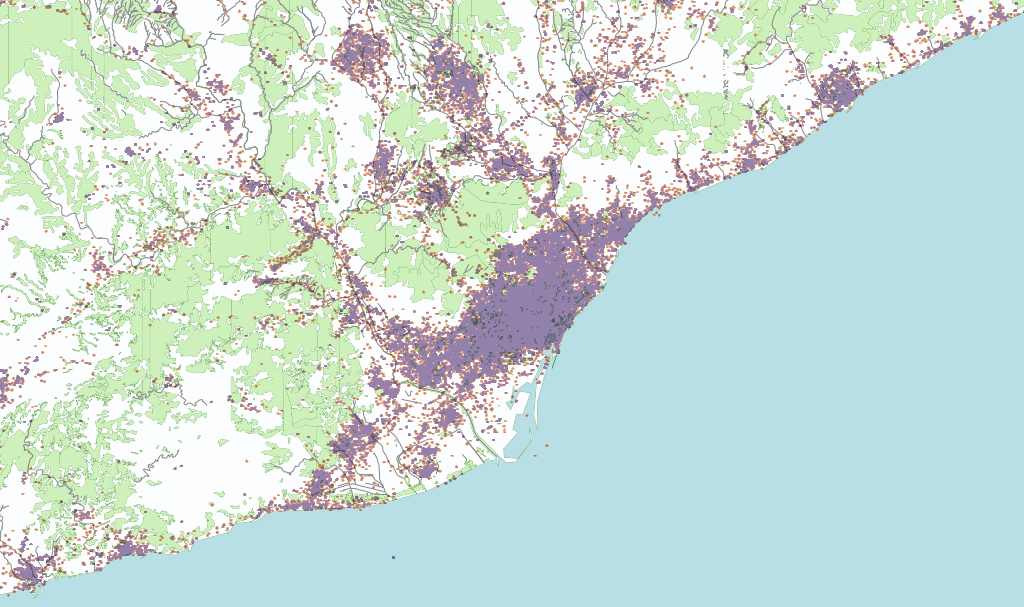}}
\caption{Geo-tweets in the area of Barcelona. \DUrole{label}{fig:bcn}}
\end{figure}\begin{figure}[]\noindent\makebox[\columnwidth][c]{\includegraphics[width=\columnwidth]{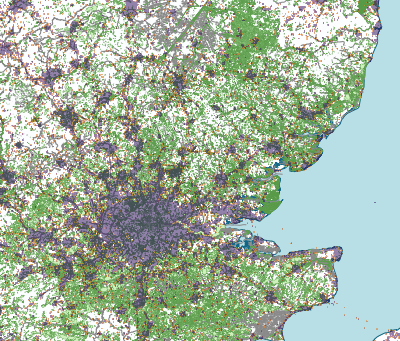}}
\caption{Geo-tweets in the area of London. \DUrole{label}{fig:lon}}
\end{figure}

Finally, in the framework of the European project EUNOIA \cite{Eunoia},  Twitter
data amongst other data will be used to caracterise and compare mobility and
location patterns in different European cities. Besides,
urban land use and transportation models will be studied by integrating the role
of the social network and new models of joint trips and joint resource use.

\section{Concluding remarks%
  \label{concluding-remarks}%
}

In summary, we have presented an example of efficient social networks data acquisition and
storage by using Python programming language and specific packages to connect
user's applications to Twitter APIs and to MongoDB distributed non relational database.

\section{Acknowledgements%
  \label{acknowledgements}%
}

Financial support from the European Comission through project FP7-ICT-2011-8 (EUNOIA),
from MINECO (Spain) and FEDER (EU) through projects FIS2007-60327 (FISICOS) and
FIS2012-30634 (\href{mailto:INTENSE@COSYP}{INTENSE@COSYP}), from CSIC through project GRID-CSIC (Ref. 200450E494) and PIE-201050E119,
and from Comunitat Auto'onoma de les Illes Balears is acknowledged.

\end{document}